\documentstyle[useAMS,epsfig]{mn2e}

\newcommand{\HI}{H\,{\sc i}}

\newcommand{\kms}{~km\,s$^{-1}$}
\newcommand{\kkms}{km\,s$^{-1}$}

\newcommand{\Msun}{$M_{\odot}$}

\title[\HI\ Mapping of Galaxies in six GEMS Groups]
      {\HI\ Mapping of Galaxies in six GEMS Groups\thanks{The 
       observations were obtained with the Australia Telescope Compact Array which 
       is funded by the Commonwealth of Australia for operations as 
       a National Facility managed by CSIRO.}}

\author[K. M. Kern et al.]
       {Katie M. Kern$^{1,2}$,
	Virginia A. Kilborn$^{1,2}$,
        Duncan A. Forbes$^{1}$ and 
	B\"arbel Koribalski$^{2}$ \\
	$^1$Centre for Astrophysics \& Supercomputing, Swinburne University of Technology,  
	    Hawthorn, VIC 3122, Australia\\
	$^2$Australia Telescope National Facility, CSIRO, 
	    P.O. Box 76, Epping, NSW 1710, Australia}

\begin{document}

\date{Received date; accepted date}

\pagerange{\pageref{firstpage}--\pageref{lastpage}} \pubyear{2006}

\maketitle

\label{firstpage}

\begin{abstract}

Here we present Australia Telescope Compact Array \HI\ maps of 16
\HI\ sources in six Group Evolution Multiwavelength Study (GEMS)
groups that were 
previously observed with the Parkes telescope. The higher spatial
resolution of the ATCA allows us to clearly identify the optical
counterparts for the first time -- 
most being associated with low surface
brightness late-type galaxies. New integrated \HI\ maps
and velocity fields for each source are presented.
We find several interacting systems; one of which contains three
galaxies within a common \HI\ envelope. 
Extended \HI\ structures in the sample are more consistent with tidal effects 
than ram pressure stripping.
We identify two \HI\ detections with previously
uncatalogued optical galaxies, and add a total of six 
newly identified group members to the NGC 3923, 5044
and 7144 groups. 

\end{abstract}

 
\section{Introduction} 

Approximately half of all galaxies reside in galaxy groups (Eke
et al.~2004).  Despite being so abundant in the universe, 
the properties of groups and their constituent
galaxies have been less well studied compared to rare 
galaxy clusters.  Most detailed multiwavelength studies to date (e.g. 
Mulchaey \& Zabludoff 1998) have been restricted to a small
number of groups. 

Galaxies in the dense environments of cluster cores 
exhibit suppressed star formation
rates, neutral hydrogen (\HI) deficiencies, bulge-dominated
morphologies, and redder colors than their field counterparts
(e.g. Wilman et al.~2005a,b). The environment of galaxy groups is
also thought to have a strong impact on the evolution of
constituent galaxies. It has been argued to be the most
effective environment for galaxy interactions and mergers due to
the low relative velocities in groups (e.g. Mulchaey \& Zabludoff
1998; Mamon 2006). Given these low velocities, and low
intragroup gas densities, it is expected that ram pressure
stripping plays a much reduced role in groups compared to
clusters, however examples do exist (e.g. Rasmussen, Ponman \&
Mulchaey 2006).

\HI\ mapping is an extremely useful technique 
in understanding galaxy groups, as most
\HI\ envelopes 
around galaxies tend to extend further out than the stellar
component, and as such, are the first component of the galaxies
to react to external forces (Broeils \& van Woerden 1994).
Therefore, the
\HI\ 
properties of group galaxies serve as a sensitive indicator
of past and present interactions, as well as internal
kinematics, mass, and orientation parameters of the
galaxies themselves.  Several studies of
\HI\ in galaxy groups have been made, showing that group galaxies 
often display \HI\ material that is extended and
stripped away from their host galaxies (e.g. Gordon et al.~2003; Kantharina et
al.~2005; Koribalski
\& Manthey 2005; Kilborn et al.~2006). In the case of compact
groups, galaxies in such environments appear to be 
deficient in \HI\ (Verdes-Montenegro et al. 2001) but see Stevens et
al. (2004) for a different view point. 

To gain a better understanding of the processes operating within
groups, the Group Evolution Multiwavelength Study
(GEMS) was established. For details of the
selection criteria and science goals, see Osmond \& Ponman~(2004)
and Forbes et al.~(2006).
Briefly, this is a multiwavelength study of 60
nearby galaxy groups, using X-ray,
optical, infrared and \HI\ data.  
All have deep ROSAT X-ray
maps available. We have also obtained wide-field  \HI\
observations of 16 southern GEMS groups using the multibeam
receiver on the Parkes Radio telescope (Kilborn et al.~2007). 
This Parkes survey, which is about twice as sensitive as
the HIPASS survey (Barnes et al. 2001), 
has discovered several new galaxy group
members (McKay et al.~2004; Kilborn et al.~2005) and interesting
cases of extended \HI\ gas (Kilborn et al.~2007). 
In the Kilborn et
al.~(2007) survey there are $\sim$50 sources for which the \HI\
detection could not be confidently assigned to a single (optical)
galaxy. This `confusion' is due to the relatively large beamsize of the
Parkes telescope, i.e. 15.5${\arcmin}$.
Here we report the first of our higher resolution (beamsize
$\sim$ 2${\arcmin}$) \HI\ maps with the 
Australian Telescope Compact Array (ATCA) of 16 confused
sources in six GEMS groups. Our aim is to identify the correct
optical counterpart and map the \HI\ at higher
spatial resolution. 



\section{Observations and Data Reduction}

Our \HI\ line observations of 16 Parkes sources were obtained
with the ATCA in `snapshot' mode over several observing runs in
2004 March, 2004 June, 2004 November, 2005 January, and 2006
January.  Data were obtained using various configurations of the
750 array, and an 8~MHz bandwidth with 512 channels.  The
observing parameters, such as the array used, date of
observations, integration time per source, phase calibrator, and
central observing frequency for the sixteen sources can be found
in Table~\ref{tab:table1}.  The primary calibrator used for all
sources was PKS 1934-638, and the secondary calibrator was
generally kept the same for individual galaxy groups.

Also listed in
Table~\ref{tab:table1} are the deconvolved beamsizes for each
field, and the noise as measured from the cleaned channel maps
near the \HI\ emission using MIRIAD task CGDISP.  
The \HI\ column density limit is calculated
using Equation~\ref{eq:colden} 
at a detection level of 3$\sigma$.
The \HI\ mass sensitivity, calculated by
Equation~\ref{eq:hisense}, is the lower
\HI\ mass limit detectable for each source, assuming a distance
D to each group from Brough et al. (2006)
and an integrated \HI\ flux of width W = 
100\kms\ at the 3$\sigma$ level (in mJy beam$^{-1}$). 
In the best case we
have a sensitivity of $\sim$ 10$^8$ M$_{\odot}$. 

\begin{equation}
\label{eq:colden}
N_{H\,{\sc i}} = \frac{1.82\times10^{18} \times 3\sigma \times 685}{beamsize \times 1.2} \,~~ atoms \, cm^{-2}
\end{equation}

\begin{equation}
\label{eq:hisense}
Sensitivity = 2.36\times10^{2} \times D^{2} \times 3\sigma \times W \,~~ M_{\odot}
\end{equation} 

The data were reduced using standard MIRIAD routines including
flagging, bandpass calibration, subtraction of the continuum, and
inverted into the real domain.  The final \HI\ cubes were made
using natural weighting and smoothed to a 
velocity resolution of 6.6 \kms.  The
channels were cleaned, giving an rms noise of $\sim$1-2 mJy
beam$^{-1}$.
The resulting \HI\
moment maps for the detections were prepared using the MIRIAD
task MOMENT, where each map was masked to the lowest \HI\
integrated intensity contour using MATHS (for most galaxies, the
lowest contour was 0.5 Jy beam$^{-1}$).

The integrated intensity maps were run through IMFIT to determine
the position of the galaxies, whereby this central position was
used in MBSPECT with the \HI\ cubes to measure the central \HI\
velocity and the width of the line profile at the 50\% peak
flux density $W_{50}$.  
To determine the total integrated \HI\ flux, the task CGDISP was
once again used on the integrated intensity maps, where multiple
polygons were drawn around the source at varying radii to
determine the mean and deviation of the integrated flux.
We note that due to the relatively short exposure times in
snapshot mode, the integrated fluxes are likely to be lower
estimates of the true total flux and we refer the reader to
Kilborn et al. (2007) for total source fluxes. Basic properties
of the 19 galaxies detected in our ATCA observations are
summarised in Table 2.

\begin{table*}
\centering
\caption{ATCA observing parameters.}
\label{tab:table1} 
\begin{tabular}{llllclllll}
\hline
GEMS Name   & Beamsize  &  N$_{\HI}$  & Sensitivity  & Noise & Obs. Freq.& Calibrator & Int. Time & Array & Date\\
 & [arcsec$^{2}$] & [$10^{18}$ cm$^{-2}$] & [$10^8$ \Msun] & [mJy
beam$^{-1}$] & [MHz] &  & [minutes] & & \\
 (1)  & (2) & (3) & (4) & (5) & (6) & (7) & (8) & (9) & (10) \\
\hline\hline
GEMS\_N1332\_03 & 150$\times$45 & 2.7 & 1.8 & 6 & 1412 & 0405-385 & 135 & 750B & 2005-01-17 \\
GEMS\_N1332\_13 & 121$\times$50 & 3.0 & 1.8 & 6 & 1412 & 0405-385 & 150 & 750B & 2005-01-17 \\
 \\	       
GEMS\_N1808\_01 & 89$\times$47  & 4.5 & 1.2 & 6 & 1412 & 0405-385 & 140 & 750B & 2005-01-17\\
 \\	       
GEMS\_N3923\_08 & 114$\times$51 & 3.1 & 1.9 & 6 & 1411 & 1151-348 & 143 & 750C & 2004-11-12 \\
GEMS\_N3923\_11 & 101$\times$58 & 2.6 & 1.6 & 5 & 1411 & 1151-348 & 156 & 750C & 2004-11-12 \\
GEMS\_N3923\_13 & 119$\times$52 & 3.0 & 1.9 & 6 & 1411 & 1151-348 & 137 & 750C & 2004-11-12 \\
 \\	       
GEMS\_N5044\_01 & 178$\times$46 & 1.9 & 3.0 & 5 & 1407 & 1245-197 & 270 & 750D & 2006-01-26 \\
GEMS\_N5044\_05 & 168$\times$51 & 2.9 & 4.8 & 8 & 1407 & 1245-197 & 132 & 750D & 2006-01-28 \\
GEMS\_N5044\_10 & 221$\times$45 & 1.9 & 3.6 & 6 & 1407 & 1245-197 & 265 & 750D & 2006-01-27 \\
GEMS\_N5044\_14 & 183$\times$51 & 2.3 & 4.2 & 7 & 1407 & 1245-197 & 132 & 750D & 2006-01-28 \\
GEMS\_N5044\_18 & 202$\times$47 & 1.9 & 3.6 & 6 & 1409 & 1245-197 & 265 & 750D & 2006-01-27 \\
 \\	       
GEMS\_N7144\_01 & 76$\times$45  & 8.1 & 3.3 & 9 & 1411 & 2134-470 & 116 & 750D & 2004-06-05 \\
GEMS\_N7144\_05 & 75$\times$50  & 2.5 & 1.1 & 3 & 1411 & 2106-413 & 588 & 750A & 2004-06-04 \\
GEMS\_N7144\_06 & 78$\times$45  & 7.0 & 2.9 & 8 & 1411 & 2134-470 & 105 & 750D & 2004-06-05 \\
GEMS\_N7144\_07 & 78$\times$52  & 2.3 & 1.1 & 3 & 1411 & 2106-413 & 639 & 750A & 2004-06-03 \\
 \\	       
GEMS\_HCG90\_11 & 90$\times$50  & 4.8 & 6.4 & 7 & 1408 & 2149-306 & 135 & 750D & 2004-06-05 \\
\hline
\end{tabular}
\flushleft The columns are (1) GEMS name (2) beamsize, 
(3) column density,
(4) sensitivity,
(5) noise, (6) central observing frequency, (7) the phase calibrator used,
(8) total integration time on the source, (9) array used, 
(10) the date of the observation.
\end{table*}

\section{Results}

We observed sixteen confused Parkes \HI\ sources in six GEMS galaxy groups
from the catalogue of Kilborn et al. (2007) and identified 19
optical counterparts to these sources  
with the superior resolution of the ATCA. 
The total \HI\ flux for these sources 
from the Parkes telescope are given in 
Kilborn et al. (2007). 
In Table~\ref{tab:hi_params} we list the ATCA \HI\ measurements
and our optical identification based on galaxies in the NASA
Extragalactic Database (NED; nedwww.ipac.caltech.edu) and the
Digital Sky Survey (DSS; www-gsss.stsci.edu/SkySurveys/SkySurveys.htm).
For three of the Parkes sources we resolved multiple \HI\ sources.
These are denoted by A or B added to the end of the 
GEMS Name of the Parkes source. In each case a catalogued 
galaxy to the A and B \HI\ source could be clearly identified.
Of the remaining 13 Parkes sources we 
were able to confidently associate 11 of them with an 
optical counterpart based on their spatial coincidence. For
two \HI\ sources (which we refer to as GEMS\_N3923\_11 and
GEMS\_N5044\_18) no previously catalogued optical galaxy could be
found in the NED database, however we could clearly identify a
small, low surface brightness galaxy in the DSS in both cases. 
For a total of six galaxies no previous redshift was catalogued. 
From the GEMS project (this work, Kilborn et al. 2007 and McKay
et al. 2004) we find all six to have locations and
redshifts consistent with group membership.  
Details of our results for individual galaxies are described in the next
section.

\begin{table*}
\caption{Properties of the detected GEMS galaxies.}
\label{tab:hi_params} 
\begin{tabular}{llrrrrrrr}
\hline
GEMS Name & Optical ID & HI position & V$_{\rm HI}$ & $W_{50}$ &
$S_{\rm HI}$ &$M_{\rm HI}$& $V_{opt}$ & Type\\
& & [$^{\rm h\,m\,s}$], [\degr\,\arcmin\,\arcsec] & [\kkms]
& [\kkms]  & [Jy \kkms] & $10^8 M_{\odot}$ & [\kkms] & \\  
(1)  & (2)        & (3)
& (4)       &  (5)    & (6) & (7) & (8) &(9) \\
\hline\hline
GEMS\_N1332\_03  & IC 1953             & 03:33:40.0,$-$21:28:39.5& 1844 & 198 & 5.9$\pm$0.1 &13.4$\pm0.9$  &1860 & SB(rs)d\\ 
GEMS\_N1332\_13  & NGC 1385            & 03:37:29.0,$-$24:30:03.6& 1492 & 207 & 17.7$\pm$0.3 &21.5$\pm1.9$ &1502 & SB(s)cd\\
\\														
GEMS\_N1808\_01  & ESO 362-G011        & 05:16:39.7,$-$37:06:09.8& 1338 & 275 & 56$\pm$1    &37.9$\pm1.5$ &1367 & Sbc\\ 
 \\					  											
GEMS\_N3923\_08  & ESO 439-G025        & 11:43:45.9,$-$30:37:13.8& 1972 & 133 & 5.4$\pm$0.1 &7.2$\pm0.6$ &---  & Sb\\  
GEMS\_N3923\_11 & ---	    & 11:50:17.2,$-$30:01:25.0 &1594 & 57  & 0.6$\pm$0.2 &1.4$\pm0.4$ & --- & ---\\  
GEMS\_N3923\_13  & [KK2000] 47         & 11:57:30.0,$-$28:07:39.8& 2112 & 35  & 1.7$\pm$0.1 &1.8$\pm0.3$ &---  & Irr/Sph\\ 
\\					  											
GEMS\_N5044\_01  & LEDA 083818         & 13:13:59.6,$-$16:47:28.1& 3037 & 140 & 1.0$\pm$0.2 &8.7$\pm1.0$ &---  &  Sd\\  
GEMS\_N5044\_05  & RC3 1303.0-1530   & 13:05:34.0,$-$15:45:09.0 &2875 & 99  & 1.7$\pm$0.1 &7.7$\pm1.0$ &---  &  SB(s)dm\\  
GEMS\_N5044\_10  & [MMB2004] J1320-1427  & 13:20:13.5,$-$14:27:36.4& 2723 & 45  & 2.4$\pm$0.1&9.7$\pm1.1$ & ---  &  --\\  
GEMS\_N5044\_14  & RC3 1305.5-1430     & 13:08:06.3,$-$14:46:43.4& 2578 & 106 & 1.5$\pm$0.5 &11.1$\pm1.1$ &---  &  IB(s)m\\  
GEMS\_N5044\_18 & ---     & 13:11:33.1,$-$14:40:39.8 & 2421 &72  & 0.3$\pm$0.1 &6.3$\pm0.9$ &---  &  --\\ 
\\					  
GEMS\_N7144\_01A & ESO 236-G039        & 21:45:13.8,$-$49:00:32.0& 1598 & 64  &   8$\pm$2   &26.2$\pm1.8$ &---  & Int.\\  
GEMS\_N7144\_01B & KTS 65 A,B          & 21:44:53.7,$-$49:00:27.7& 1586 & 81  &  12$\pm$2   &26.2$\pm1.8$ &---  & Int.\\  
GEMS\_N7144\_05  & B215242.56-492853.8$^*$        & 21:55:56.8,$-$49:14:37.3& 1849  & 81  & 2.8$\pm$0.1 &5.9$\pm0.8$ & --- & ---\\  
GEMS\_N7144\_06A & ESO 236-G036   & 21:42:52.9,$-$47:51:22.8 &2004  & 77  & 1.9$\pm$0.1 &24.8$\pm1.5$ & ---  & IB(s)m\\  
GEMS\_N7144\_06B & ESO 236-G035    & 21:42:48.3,$-$47:59:05.1 &1958  & 168 & 2.5$\pm$0.1 &24.8$\pm1.5$ &2086 & SB(rs)c\\  
GEMS\_N7144\_07  & B213743.74-4654387$^*$         & 21:41:07.1,$-$46:38:50.1& 1939  & 31  & 0.7$\pm$0.1 &3.0$\pm0.57$ & --- & ---\\
\\																
GEMS\_HCG90\_11A & NGC 7204       & 22:06:53.2,$-$31:03:07.1 & 2542& 178 & 2.5$\pm$0.1 &23.0$\pm2.3$ & 2630 & S0/a pec.\\  
GEMS\_HCG90\_11B & ESO 467-G002   & 22:06:09.9,$-$31:05:19.0 & 2583& 40  & 1.6$\pm$0.1 &23.0$\pm2.3$ & 2550 & Sd\\ 
\hline
\end{tabular}
\flushleft The columns are: 
(1) GEMS name, 
(2) Optical identification ($^*$ refers to the APMUKS(BJ) catalogue),
(3) fitted \HI\ center position,
(4) observed \HI\ systemic velocity,
(5) 50\% velocity width, 
(6) measured ATCA \HI\ flux, 
(7) measured Parkes \HI\ mass (from Kilborn et al. 2007).
(8) optical velocity from NED, 
(9) morphological type from NED 
\end{table*}


\subsection{NGC 1332 Group}

{\bf IC 1953} This is a late-type barred spiral galaxy. 
The \HI\ detected here does not 
extend further than the stellar component, and has peaks to
either side of the nucleus 
(Figure~\ref{fig:hispectra1}).  
The galaxy to the north-east of IC 1953 (in the DSS image) 
is ESO 548-G040 which
lies in the background at a velocity of 4061 \kms.\\

\noindent
{\bf NGC 1385} This is a barred spiral galaxy.  
The \HI\ of this
galaxy is evenly distributed and extends beyond the stellar
component 
(Figure~\ref{fig:hispectra1}). The \HI\ velocity contours show a
normal differentially rotating disk, although a slight warp appears in the
outer regions.  The \HI\ spectrum is fairly symmetric, showing one
central peak.

\subsection{NGC 1808 Group}

\begin{figure*}
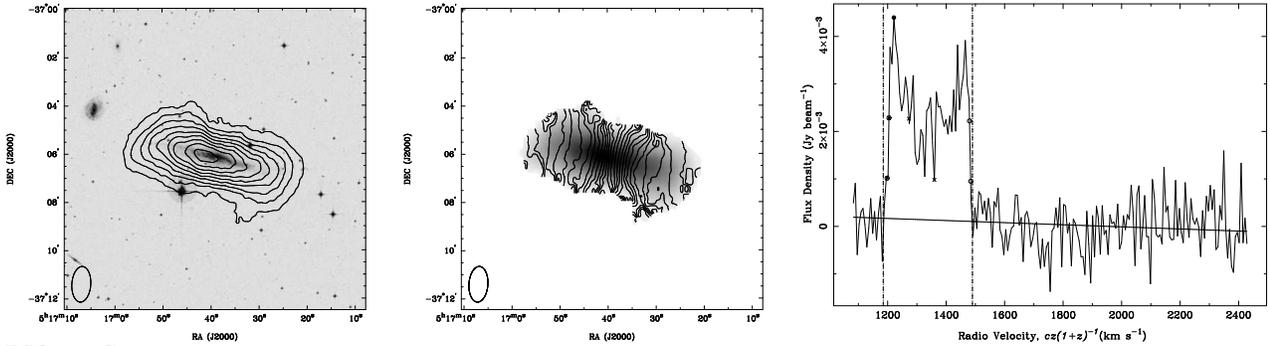
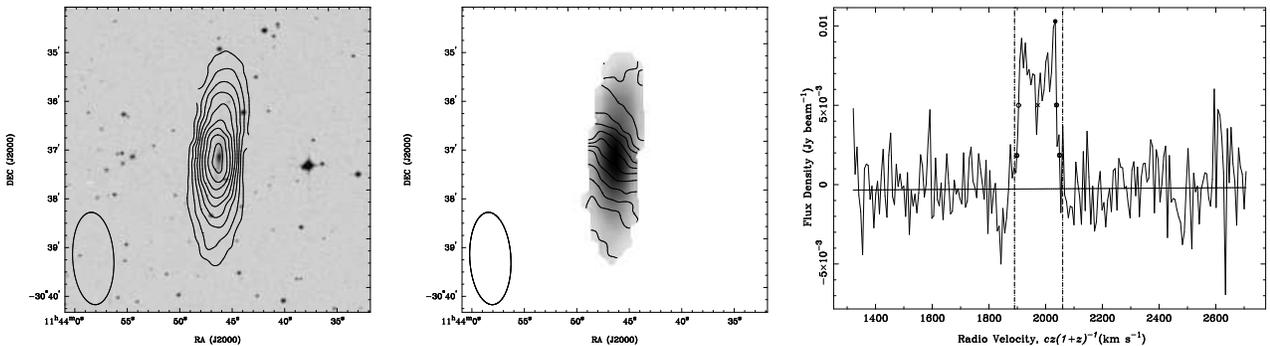

\begin{tabular}{lll}
\mbox{IC 1953} &  & \\
 \mbox{\psfig{file=n1332_07.overlay,width=4.5cm,angle=-90}}&
 \mbox{\psfig{file=n1332_07.cont,width=4.5cm,angle=-90}}&
 \mbox{\psfig{file=n1332_07.spectra,width=4.5cm,angle=-90}}\\
\mbox{NGC 1385} &  &  \\
 \mbox{\psfig{file=n1332_01.overlay,width=4.5cm,angle=-90}}&
 \mbox{\psfig{file=n1332_01.cont,width=4.5cm,angle=-90}}&
 \mbox{\psfig{file=n1332_01.spectra,width=4.5cm,angle=-90}}\\  
\mbox{ESO 362-G011} &  &  \\
 \mbox{\psfig{file=n1792_04.overlay,width=4.5cm,angle=-90}}&
 \mbox{\psfig{file=n1792_04.cont,width=4.5cm,angle=-90}}&
 \mbox{\psfig{file=n1792_04.spectra,width=4.5cm,angle=-90}}\\
\mbox{ESO 439-G025} &  &  \\
 \mbox{\psfig{file=n3923_01.overlay,width=4.5cm,angle=-90}}&
 \mbox{\psfig{file=n3923_01.cont,width=4.5cm,angle=-90}}&
 \mbox{\psfig{file=n3923_01.spectra,width=4.5cm,angle=-90}}\\
\end{tabular}
\caption{\HI\ moment and spectral line maps of IC 1953, NGC 1385,
ESO 362-G011, and ESO 439-G025, respectively, according to row.
Column one shows the integrated \HI\ intensity (contours)
overlaid on an optical DSS image. The contour levels begin at 0.5
Jy~beam$^{-1}$\kms, and increment by 1.0 Jy~beam$^{-1}$\kms\ for
NGC 1385 and ESO 362-G011, and increment by 0.5
Jy~beam$^{-1}$\kms\ for IC 1953 and ESO 439-G025. The second
column shows the velocity contours over the \HI\ integrated
intensity greyscale, where the maps were masked below the 0.5
Jy~beam$^{-1}$\kms\ level.  The velocity increments are 10\kms,
with starting velocities at 1770, 1425, 1200, and 1920\kms,
respectively.  The beam size is shown in the lower left of the
\HI\ line maps. 
Column three gives the \HI\ spectra.  The fitted baseline is
shown, and the \HI\ peak flux density is marked with a filled
circle. The $W_{20}$ and $W_{50}$ velocity widths are shown by
the open circles (outer fit), and crosses (inner fit). The
velocity region between the vertical lines in the spectra were
disregarded in the baseline fit.}
\label{fig:hispectra1}
\end{figure*}

{\bf ESO 362-G011} This is a highly inclined Sbc spiral.  The
integrated \HI\ is evenly distributed, following the position
angle (P.A.) of the
stellar component at 76$^\circ$, and extends above and below the
disk (Figure~\ref{fig:hispectra1}).  There is also a slight
extension of the
\HI\ distribution to the north-east and south-west.  
The velocity field reveals an evenly rotating galactic
disk.  The galaxy seen to the left of ESO 362-G011 (on the DSS image) is a background galaxy ESO 362-G012 at a velocity of 4638 \kms.
The \HI\ spectrum follows the classic double horned profile. 

%
%

\subsection{NGC 3923 Group}

{\bf ESO 439-G025} This edge-on Sb spiral has a bright central
nucleus  (Figure~\ref{fig:hispectra1}).  
The \HI\ follows the stellar component with a uniform 
distribution.  
The
velocity field is regular, and shows a slight change in the P.A.
near the very edges of the \HI\ emission.  The \HI\ spectrum is
fairly symmetric. Although NED lists no optical redshift, an \HI\
velocity was known previously from HIPASS (i.e. 1986 \kms). The
Parkes \HI\ source is resolved in our ATCA observations and is associated with
the optical galaxy ESO 439-G025, which can now be confirmed as a 
member of the NGC 3923 group.\\

\begin{figure*}
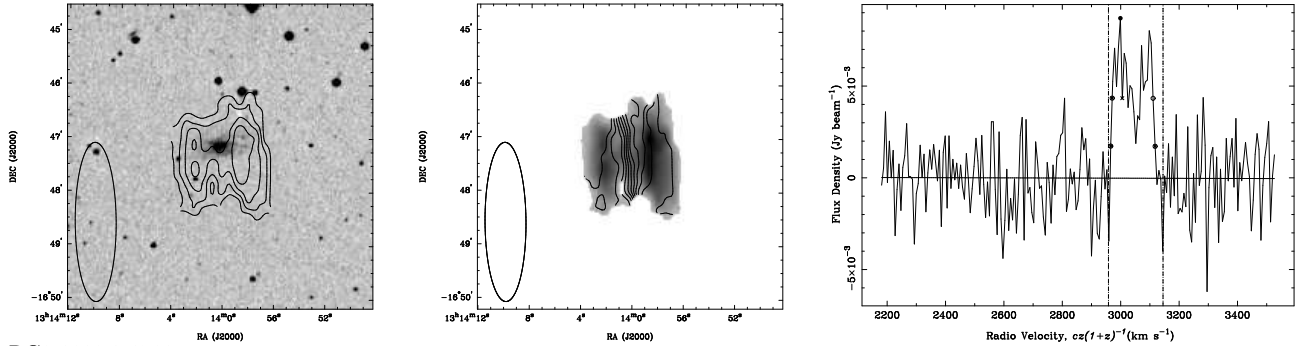
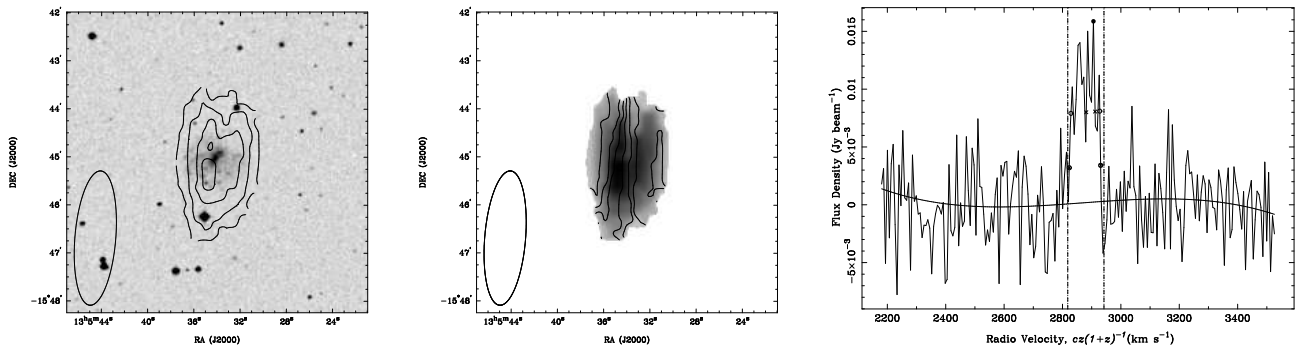

\begin{tabular}{lll}
\mbox{NGC 3923-11} &  &  \\
 \mbox{\psfig{file=n3923_13.overlay,width=4.5cm,angle=-90}}&
 \mbox{\psfig{file=n3923_13.cont,width=4.5cm,angle=-90}}&
 \mbox{\psfig{file=n3923_13.spectra,width=4.5cm,angle=-90}}\\ 
\mbox{[KK2000] 47} &  &  \\
 \mbox{\psfig{file=n3923_11.overlay,width=4.5cm,angle=-90}}&
 \mbox{\psfig{file=n3923_11.cont,width=4.5cm,angle=-90}}&
 \mbox{\psfig{file=n3923_11.spectra,width=4.5cm,angle=-90}}\\
\mbox{LEDA 083818} &  &  \\
 \mbox{\psfig{file=n5044_1.overlay,width=4.5cm,angle=-90}}&
 \mbox{\psfig{file=n5044_1.omo1,width=4.5cm,angle=-90}}&
 \mbox{\psfig{file=n5044_1.spectra,width=4.5cm,angle=-90}}\\ 
\mbox{RC3 1303.0-1530} &  &  \\
 \mbox{\psfig{file=n5044_5.overlay,width=4.5cm,angle=-90}}&
 \mbox{\psfig{file=n5044_5.omo1,width=4.5cm,angle=-90}}&
 \mbox{\psfig{file=n5044_5.spectra,width=4.5cm,angle=-90}}\\ 
\end{tabular}
\caption{\HI\ moment and spectral line maps of GEMS\_N3923\_11,
[KK2000] 47, LEDA 083818, and RC3 1303.0-1530 respectivly,
according to row.  Column one shows the integrated \HI\ intensity
(contours) overlaid on an optical DSS image.  The contour levels
for GEMS\_N3923\_11 and LEDA 083818, start at 0.3 Jy~beam$^{-1}$\kms,
with increments of 0.2 Jy~beam$^{-1}$\kms.  For all other
galaxies, the contour levels begin at, and increment by, 0.5
Jy~beam$^{-1}$\kms.  The beam is shown in the lower left of each
figure.  Column two gives the velocity contours over the \HI\
integrated intensity greyscale, where the maps were masked below
the 0.5 Jy~beam$^{-1}$\kms\ level, except for GEMS\_N3923\_11 and LEDA
083818, where the maps were masked below 0.3 Jy~beam$^{-1}$\kms.
The velocity increment for [KK2000] 47 is 2\kms, where the rest
have velocity increments of 10\kms.  Starting velocities are
2990, 2850, 1575, 2110\kms, respectively. 
The beam size is shown in the lower left of the
\HI\ line maps. The third column shows
the \HI\ spectra. The spectrum of GEMS\_N3923\_11 was Hanning
smoothed to a velocity resolution of 13.2\kms. The fitted
baseline is shown and the \HI\ peak flux density is marked with a
filled circle. The $W_{20}$ and $W_{50}$ velocity widths are
shown by the open circles (outer fit), and crosses (inner
fit). The velocity region between the vertical lines in the
spectra were disregarded in the baseline fit.}
\label{fig:hispectra2}
\end{figure*}

\noindent
{\bf GEMS\_N3923\_11} is a previously uncatalogued low surface
brightness, irregular galaxy
as seen on the DSS (Figure~\ref{fig:hispectra2}). 
Its \HI\ velocity and spatial location make
it a new member of the NGC 3923 group. The \HI\ is
unresolved with an   
\HI\ spectrum that shows a single central peak.\\

\noindent
{\bf [KK2000] 47} has been previously catalogued by Karachentseva
\& Karachentsev (2000), who classified it as an Irr/Sph.
There is no obvious nucleus nor spiral arms in the stellar
component (Figure~\ref{fig:hispectra2}).  
The galaxy appears very diffuse, with low optical surface
brightness.  The \HI\ is unresolved.
The \HI\ spectrum shows a central
peak. Although NED lists no optical redshift, an \HI\
velocity of 2125 $\pm$ 2 \kms\ was measured previously by Karachentseva
\& Karachentsev (2000) which
places it in the NGC 3923 group.\\

\subsection{NGC 5044 Group}

{\bf LEDA 083818} 
This galaxy is classified as a late-type spiral. 
The \HI\ distribution peaks on both sides of the
galaxy (Figure~\ref{fig:hispectra2}). 
The \HI\ spectrum is a symmetrical double horned profile. The
GEMS redshift confirms it as a new member of the NGC 5044 group. \\

\noindent
{\bf RC3 1303.0-1530} This galaxy is classified as 
a barred dwarf spiral. The \HI\ has a regular distribution which is
nearly centered on the bright nucleus of the galaxy 
(Figure~\ref{fig:hispectra2}).
The velocity contours reveal a rotating disk,
and the \HI\ spectrum shows a central peak. The
GEMS redshift confirms it as a new member of the NGC 5044 group.\\

\begin{figure*}
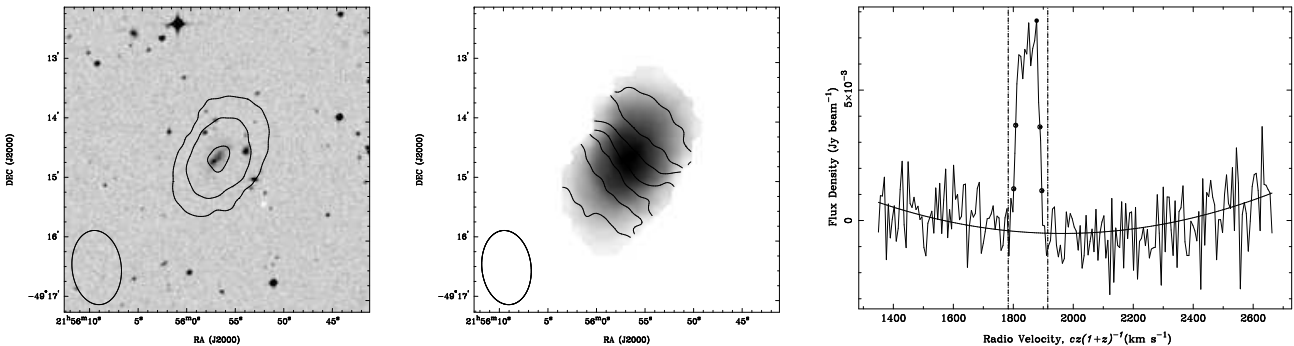

\begin{tabular}{lll}
\mbox{[MMB2004]J1320-1427} &  &  \\
 \mbox{\psfig{file=n5044_10.overlay,width=4.5cm,angle=-90}}&
 \mbox{\psfig{file=n5044_10.omo1,width=4.5cm,angle=-90}}&
 \mbox{\psfig{file=n5044_10.spectra,width=4.5cm,angle=-90}}\\
\mbox{[RC3] 1305.5-1430} &  &  \\
 \mbox{\psfig{file=n5044_14.overlay,width=4.5cm,angle=-90}}&
 \mbox{\psfig{file=n5044_14.omo1,width=4.5cm,angle=-90}}&
 \mbox{\psfig{file=n5044_14.spectra,width=4.5cm,angle=-90}}\\
\mbox{GEMS\_5044\_18} &  &  \\
 \mbox{\psfig{file=n5044_18.overlay,width=4.5cm,angle=-90}}&
 \mbox{\psfig{file=n5044_18.omo1,width=4.5cm,angle=-90}}&
 \mbox{\psfig{file=n5044_18.spectra,width=4.5cm,angle=-90}}\\
\mbox{B215242.56-492853.8} &  &  \\
 \mbox{\psfig{file=n7144_3.overlay,width=4.5cm,angle=-90}}&
 \mbox{\psfig{file=n7144_3.omo1,width=4.5cm,angle=-90}}&
 \mbox{\psfig{file=n7144_3.spectra,width=4.5cm,angle=-90}}\\ 
\end{tabular}
\caption{\HI\ moment and spectral line maps of [MMB2004] J1320-1427, RC3
1305.5-1430, GEMS\_N5044\_18, and B215242.56-492853.8,
respectivly, according to row.  Column one shows the integrated
\HI\ intensity (contours) overlaid on an optical DSS image.  For
GEMS\_N5044\_18, the contour levels begin at 0.28
Jy~beam$^{-1}$\kms, and increment by 0.04 Jy~beam$^{-1}$\kms.
For all others, the contour levels begin at, and increment by,
0.5 Jy~beam$^{-1}$\kms. The second column shows the velocity
contours over the \HI\ integrated intensity greyscale, where the
maps were masked below the first contour mentioned above.  The
velocity increments are 10\kms, with starting velocities at 2718,
2540, 2402, 1817\kms, respectively.  
The beam size is shown in the lower left of the
\HI\ line maps. Column three gives the \HI\
spectra.  The fitted baseline is shown, and the \HI\ peak flux
density is marked with a filled circle. The $W_{20}$ and $W_{50}$
velocity widths are shown by the open circles (outer fit), and
crosses (inner fit). The velocity region between the vertical
lines in the spectra were disregarded in the baseline fit.}
\label{fig:hispectra3}
\end{figure*}

\noindent
{\bf [MMB2004] J1320-1427} has been previously studied as part of
our GEMS project by McKay
et al.  (2004), with the 750A array on the ATCA for a full 12
hour integration.  The integrated \HI\ flux as detected by McKay et
al.  (2004) is similar to the Parkes measurements (Kilborn et
al.~2007). 
Our ATCA observations, taken in
snapshot mode, did not recover the entire \HI\ flux, but the
spectra and corresponding \HI\ integrated intensity map
(Figure~\ref{fig:hispectra3}) are consistent with 
McKay et al. (2004).  The spectrum is
centrally peaked, and the integrated \HI\ intensity shows an
extension to the south-east, with most of the \HI\ centered on
two bright stellar components.
McKay et al
(2004) show that a deep R-band image reveals a faint halo around the two
bluer and brighter regions visible in the DSS image.\\

\noindent
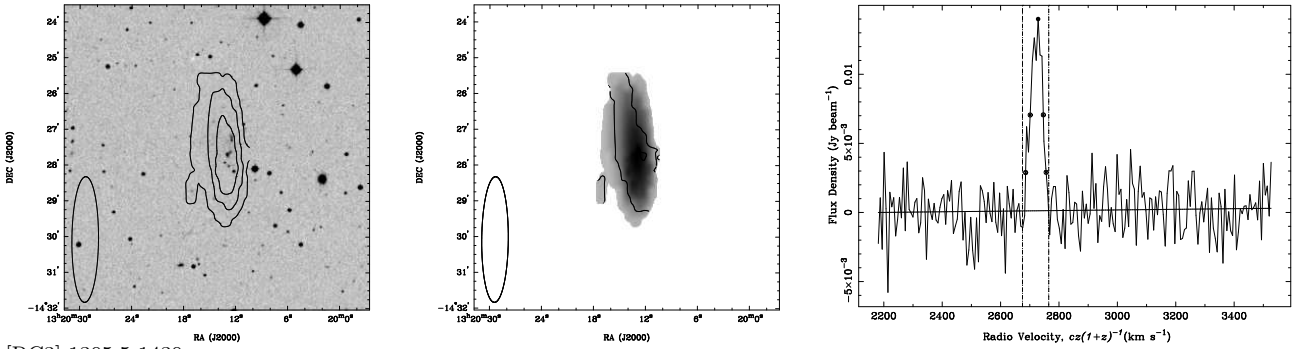
{\bf RC3 1305.5-1430} The stellar
component shows a distinct, but small nucleus, with an irregular
shape (Figure~\ref{fig:hispectra3}).  
The \HI\ component shows a peak at the optical nucleus,
where the \HI\ extends further south. 
The velocity field is unresolved, and the \HI\ spectrum
is asymmetric with more \HI\ on the approaching side. The
GEMS redshift confirms it as a new member of the NGC 5044 group.\\

\noindent
{\bf GEMS\_N5044\_18} is a previously uncatalogued very faint low
surface brightness galaxy (Figure~\ref{fig:hispectra3}).
Its \HI\ velocity and spatial location make it a new member of
the NGC 5044 group. 
The stellar component is sufficiently 
faint on the DSS  that it can only be
classified as an irregular galaxy.  The \HI\ is unresolved 
and the \HI\ spectrum shows a single central peak. \\

\subsection{NGC 7144 Group}

\begin{figure*}
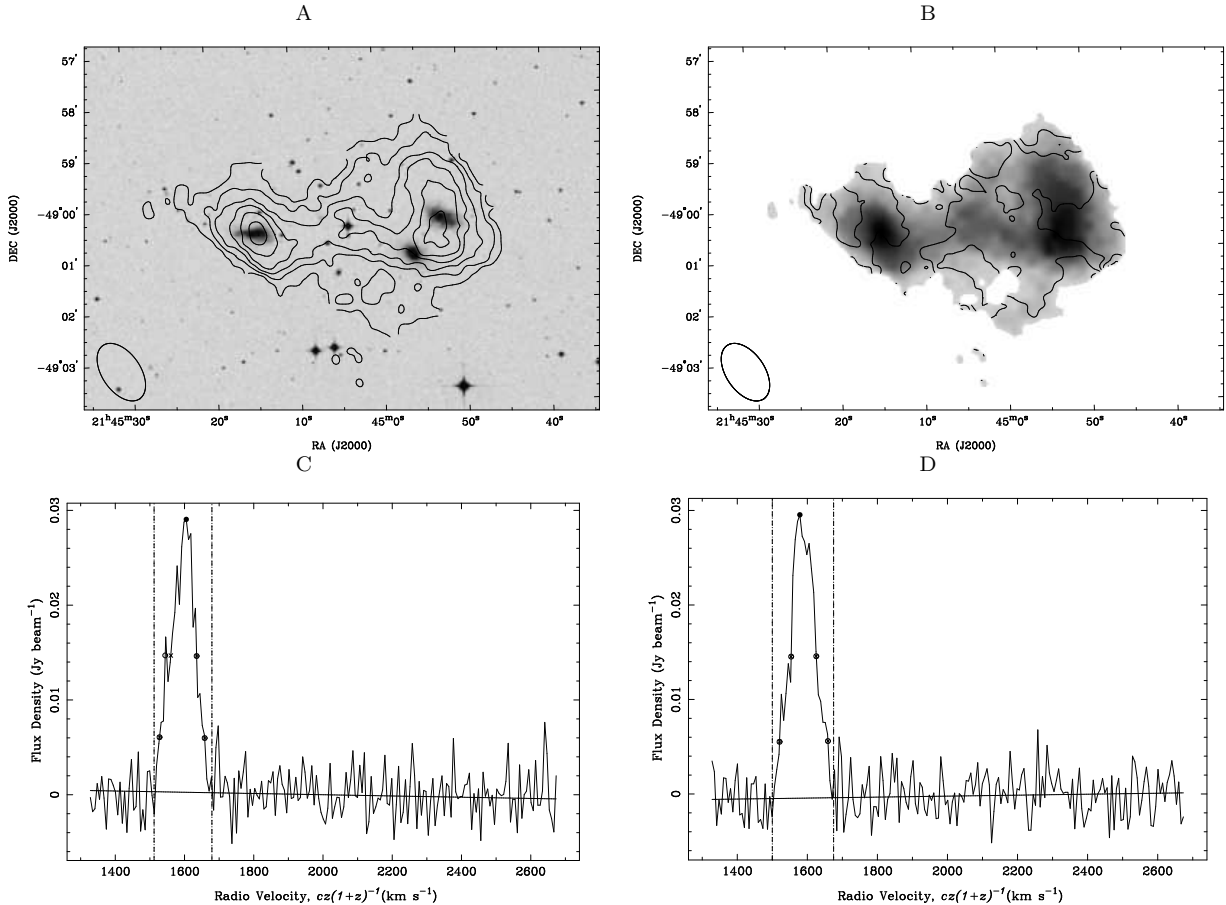

\begin{tabular}{cc}
\mbox{A} & \mbox{B} \\
 \mbox{\psfig{file=n7144-04.overlay,width=5.4cm,angle=-90}}&
 \mbox{\psfig{file=n7144_04.omo1,width=5.4cm,angle=-90}}\\ 
\mbox{C} & \mbox{D} \\
 \mbox{\psfig{file=n7144-04a.spectra,width=5.4cm,angle=-90}}&
 \mbox{\psfig{file=n7144-04b.spectra,width=5.4cm,angle=-90}}\\ 
\end{tabular}
\caption{\HI\ moment and spectral line maps of ESO 236-G039, and
KTS 65 A\&B.  (A) The integrated \HI\ intensity (contours)
overlaid on an optical DSS image. The contour levels begin at 0.3
Jy~beam$^{-1}$\kms, and increment by 0.2 Jy~beam$^{-1}$\kms.  (B)
The velocity contours over the \HI\ integrated intensity
greyscale, where the map was masked below the 0.3
Jy~beam$^{-1}$\kms\ level.  The levels start at 1582\kms, and
increments by 8\kms\ to 1606\kms.  
The beam size is shown in the lower left of panels A and B. 
(C) The \HI\ spectrum of ESO
236-G039.  (D) The \HI\ spectrum of KTS 65 A\&B.  Both spectra
were Hanning smoothed to a velocity resolution of 13.2\kms.  The
fitted baseline is shown, and the \HI\ peak flux density is
marked with a filled circle. The $W_{20}$ and $W_{50}$ velocity
widths are shown by the open circles (outer fit), and crosses
(inner fit). The velocity region between the vertical lines in
the spectra were disregarded in the baseline fit.}
\label{fig:main1}
\end{figure*}

\begin{figure*}
 \psfig{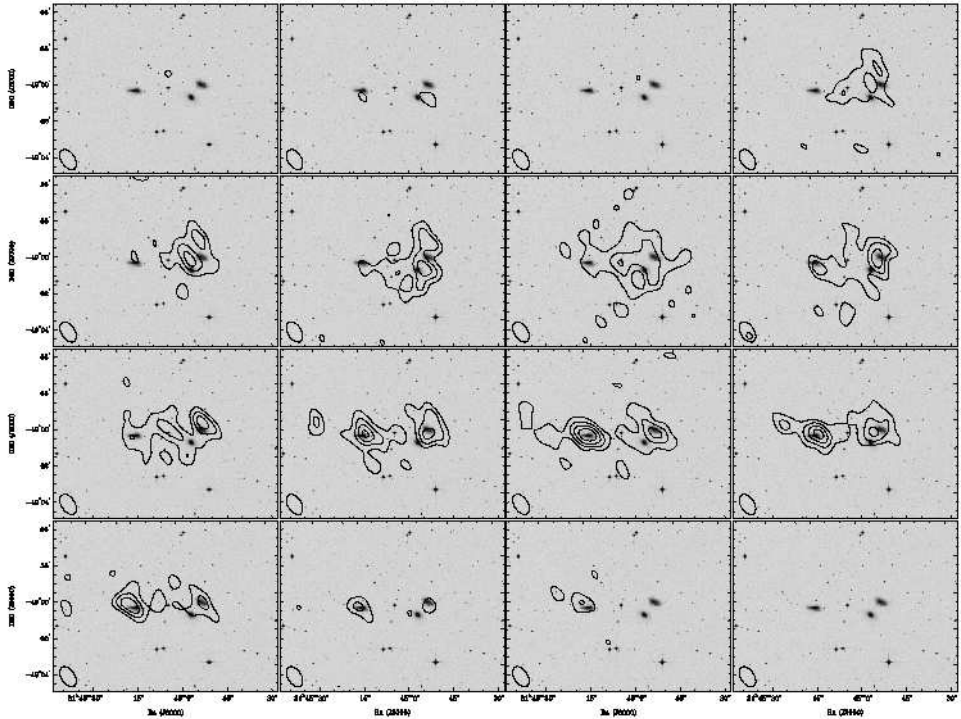}
\caption{Channel maps of ESO 236-G039 (left) and KTS 65 A\&B (right), overlaid
on a DSS map.  The panels step in velocity from left to right, then
top to bottom. The contours begin at 0.028 Jy~beam$^{-1}$\kms,
and increment by 0.01 Jy~beam$^{-1}$\kms. The beamsize is shown
lower left. }
\label{fig:main0}
\end{figure*}

{\bf ESO 236-G039} and {\bf KTS 65 A,B} 
This system consists of three irregular galaxies that are interacting. 
In Figure~\ref{fig:main1} our integrated \HI\ map
reveals a common \HI\ envelope with peaks centred on ESO 236-G039 and between
the two nuclei of KTS
65A,B with a bridge of \HI\ connecting the three galaxies. The \HI\
spectrum of ESO 236-G039 and KTS 65 A,B are both symmetric. 
The \HI\ velocity map is irregular with some
velocity substructure. The \HI\ channel maps
(Figure~\ref{fig:main0}) show that the lower
velocity \HI\ gas is peaked between KTS 65 A,B, it extends
smoothly in a bridge towards ESO 236-G039, and ends with the
highest velocity gas peaked on ESO 236-G039. Given the projected
separation of 2.75 kpc and a \HI\ velocity difference of 
$\le$12 \kms, the crossing time is $\sim$ 100 Myr. This suggests that the
three galaxies will merge into a single merger-remnant galaxy on
a short timescale. The fate of the \HI\ gas will depend on how
efficiently it is converted into stars, ionised or expelled from
the system by supernovae. \\

\noindent
{\bf [APMUKS(BJ)] B215242.56-492853.8} This is a 
small, compact galaxy. The stellar component
is slightly elongated along the same P.A. as the integrated \HI\
(Figure~\ref{fig:hispectra3}). 
The \HI\ is evenly distributed, extending much further out than
the stellar counterpart. The velocity field 
shows regular rotation. The \HI\ spectrum has a single
central peak. Although NED lists no optical redshift, an \HI\
velocity was known previously from the HIPASS survey (i.e. 1865
\kms) which is confirmed by our ATCA observations and hence we
can assign it to the NGC 7144 group.\\
\\

\begin{figure*}
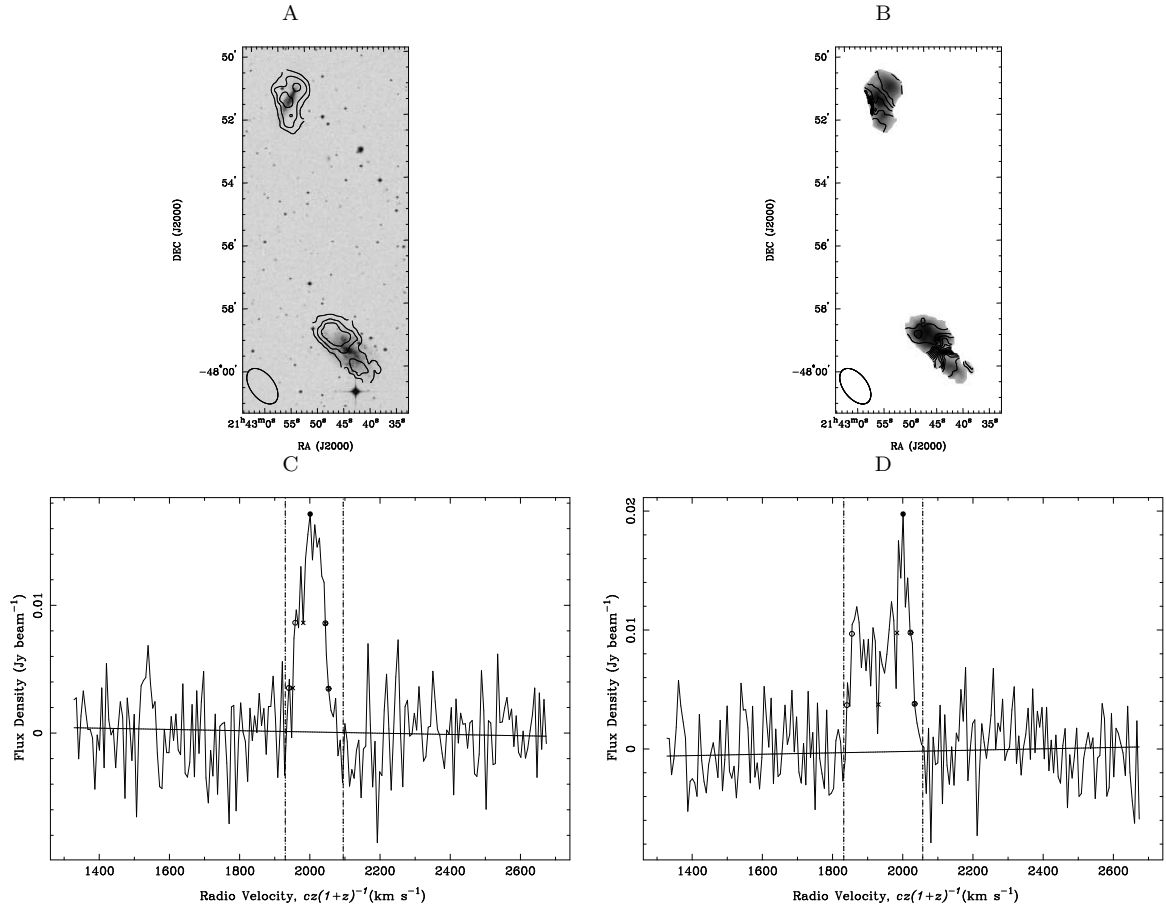

\begin{tabular}{cc}
\mbox{A} & \mbox{B} \\
 \mbox{\psfig{file=n7144_05c.overlay,width=5.4cm,angle=-90}}&
 \mbox{\psfig{file=n7144_05c.omo1,width=5.4cm,angle=-90}} \\
\mbox{C} & \mbox{D} \\
 \mbox{\psfig{file=n7144-05a.spectra,width=5.4cm,angle=-90}} &
 \mbox{\psfig{file=n7144-05b.spectra,width=5.4cm,angle=-90}}\\
\end{tabular}
\caption{\HI\ moment and spectral line maps of ESO 236-G036 and
ESO 236-G035.  ESO 236-G036 and ESO 236-G035 are mapped together
due to their small angular separation, where ESO 236-G036 is at
the top, and ESO 236-G035 is on the bottom. (A) The integrated
\HI\ intensity (contours) overlaid on an optical DSS image.  The
contour levels begin at, and increment by, 0.5
Jy~beam$^{-1}$\kms.  (B) The velocity contours over the \HI\
integrated intensity greyscale, where the maps were masked below
the 0.5 Jy~beam$^{-1}$\kms\ level.  The velocity increments are
10\kms, with a starting velocity of 1865\kms. 
The beam size is shown in the lower left of panels A and B. 
(C) The \HI\
spectra of ESO 236-G036. (D) The \HI\ spectra for ESO 236-G035.
The fitted baseline is shown, and the \HI\ peak flux density is
marked with a filled circle. The $W_{20}$ and $W_{50}$ velocity
widths are shown by the open circles (outer fit), and crosses
(inner fit). The velocity region between the vertical lines in
the spectra were disregarded in the baseline fit.}
\label{fig:hispectra5}
\end{figure*}

\noindent
{\bf ESO 236-G036} is in an interacting pair with ESO 236-G035.
The \HI\ follows the stellar component in the north but 
extends beyond the optical galaxy to the south-west
(Figure~\ref{fig:hispectra5}). This extended \HI\ may be due to 
the interaction
with ESO 236-G035.  The \HI\ map reveals several
peaks.
The velocity contours are
regular
and the \HI\ spectrum is centrally peaked. We
also note that the combined ATCA flux (3.4 Jy \kms) is less
than that for the Parkes flux GEMS\_N7144\_6 (5.6 Jy \kms)
suggesting additional \HI\ may lie between the two galaxies. 
At a distance of 22.8 Mpc,
this galaxy is separated by 52 kpc in projection from ESO
236-G035, and has a difference in velocity of 46 \kkms.\\

\noindent
{\bf ESO 236-G035} 
The \HI\ distribution 
peaks at either side of the nucleus
(Figure~\ref{fig:hispectra5}).
The \HI\ distribution shows a small extension to
the south-west, perhaps due to an interaction with ESO 236-G036. 
The \HI\ spectrum is asymmetric
with most of the \HI\ in the receeding side, which is closer to
ESO 236-G036. \\

\begin{figure*}
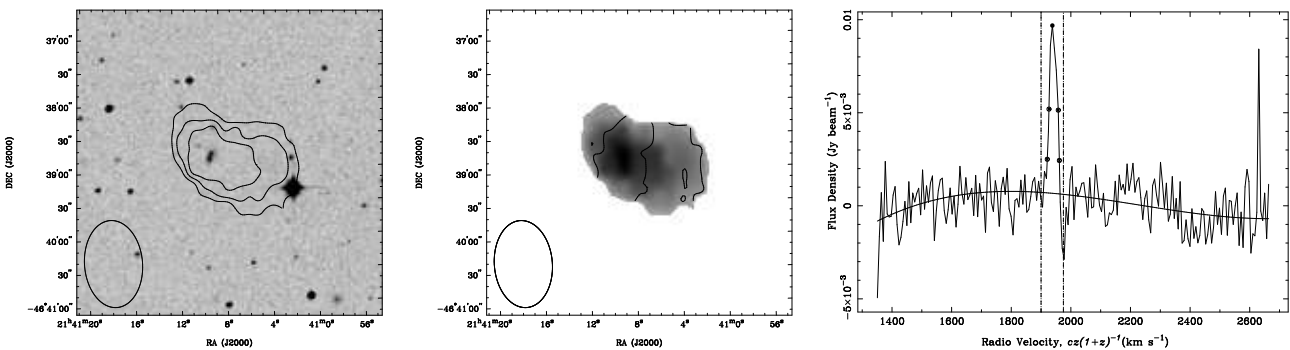

\begin{tabular}{lll}
\mbox{} &  &  \\
 \mbox{\psfig{file=n7144_7.overlay,width=4.5cm,angle=-90}}&
 \mbox{\psfig{file=n7144_7.omo1,width=4.5cm,angle=-90}}&
 \mbox{\psfig{file=n7144_7.spectra,width=4.5cm,angle=-90}}\\
\end{tabular}
\caption{\HI\ moment and spectral line maps of B213743.74-4654387.
Column one gives the integrated \HI\ intensity (contours)
overlaid on an optical DSS image.  The contour levels begin at
0.3 Jy~beam$^{-1}$\kms\ and increment by 0.1 Jy~beam$^{-1}$\kms.
Column two gives the the velocity contours over the \HI\
integrated intensity greyscale, where the maps were masked below
the 0.3 Jy~beam$^{-1}$\kms\ level.  The velocity increments are
10\kms, with a starting velocity of 1934\kms.  
The beam size is shown in the lower left of the
\HI\ line maps. Column 3 gives the
\HI\ spectrum.  The fitted baseline is shown, and the \HI\ peak flux density is marked with a filled circle. The $W_{20}$ and $W_{50}$ velocity widths are shown by the open circles (outer fit), and crosses (inner fit). The velocity region between the vertical lines in the spectra were disregarded in the baseline fit.}
\label{fig:hispectra4}
\end{figure*}

\noindent
{\bf [APMUKS(BJ)] B213743.74-4654387} This is a small, compact
galaxy (Figure~\ref{fig:hispectra4}).
The stellar component lies entirely within the
highest contour of the \HI\, where the \HI\ extends further to the
west, and has an oblong shape.  The velocity contours
are regular across the disk, and the \HI\ spectrum has a single central
peak. The
GEMS redshift confirms it as a new member of the NGC 7144 group.\\

\subsection{HCG 90 Group}

\begin{figure*}
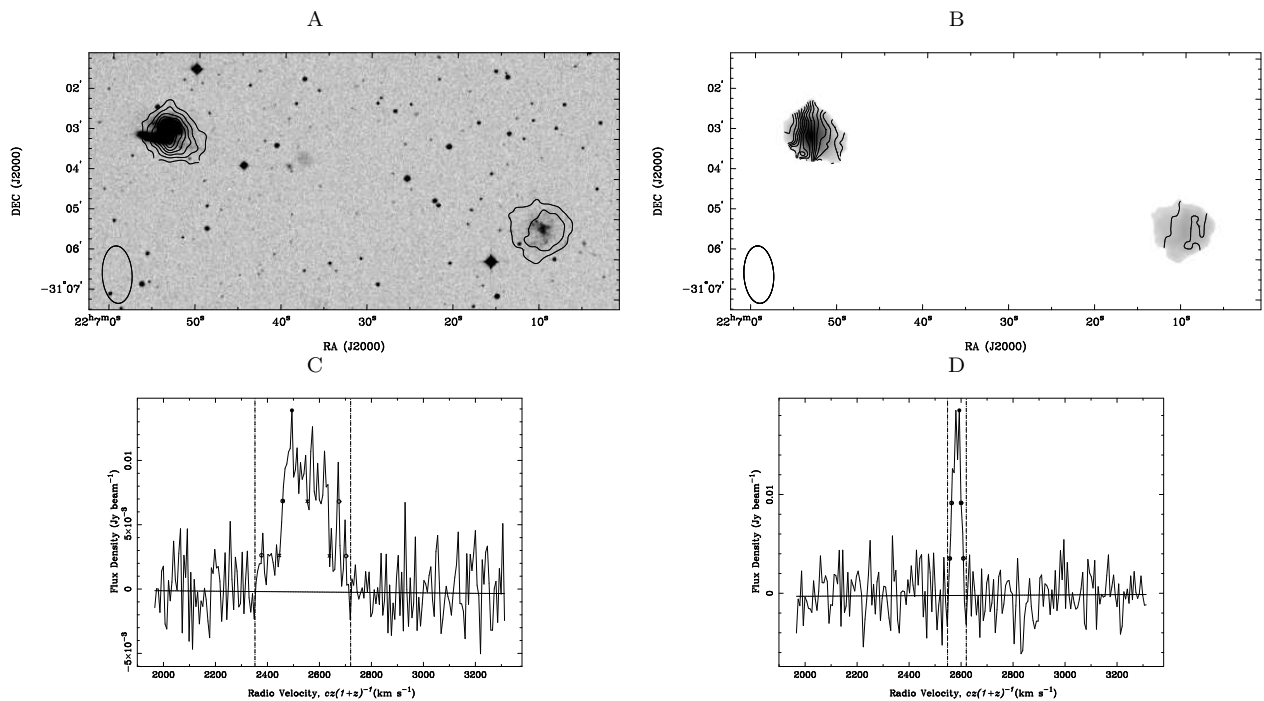

\begin{tabular}{cc}
\mbox{A} & \mbox{B} \\
 \mbox{\psfig{file=n7176-09c.overlay,width=4.0cm,angle=-90}} &
 \mbox{\psfig{file=n7176-09c.omo1,width=4.0cm,angle=-90}} \\
\mbox{C} & \mbox{D} \\
 \mbox{\psfig{file=n7176-09a.spectra,width=4.0cm, angle=-90}} &
 \mbox{\psfig{file=n7176-09b.spectra,width=4.0cm,angle=-90}}\\
\end{tabular}
\caption{\HI\ moment and spectral line maps of NGC 7204 and ESO
467-G002, where NGC 7204 is to the left of the maps, and ESO
467-G002 is to the right.  (A) The integrated \HI\ intensity
(contours) overlaid on an optical DSS image.  The contour levels
begin at, and increment by, 0.5 Jy~beam$^{-1}$\kms.  (B) Shows
the velocity contours overlaid on the \HI\ integrated intensity
greyscale.  The velocity increments are 10\kms, with starting
velocity of 2480 \kms.  
The beam size is shown in the lower left of panels A and B. 
(C) The \HI\ spectra of NGC 7204. (D)
gives the \HI\ spectra of ESO 467-G002.  The fitted baseline is
shown, and the \HI\ peak flux density is marked with a filled
circle. The $W_{20}$ and $W_{50}$ velocity widths are shown by
the open circles (outer fit), and crosses (inner fit). The
velocity region between the vertical lines in the spectra were
disregarded in the baseline fit.}
\label{fig:hispectra6}
\end{figure*}

{\bf NGC 7204} This galaxy is actually a strongly interacting
pair of galaxies.  The \HI\ peak is centered between the two
galaxies (Figure~\ref{fig:hispectra6}). 
The eastern side of the \HI\ does not extend past the stellar
component, while the western side 
extends further out. The \HI\ velocity field reveals a
surprisingly regular rotation.
NGC 7204 lies in a chain of galaxies with NGC 7201 and NGC
7203. Rubin (1974) present long-exposure image-tube photographs
(shown in their paper in Figure 2) of the galaxy chain. The Rubin
image of NGC 7204 suggests a complex star formation and/or dust
distribution.\\

\noindent
{\bf ESO 467-G002} This is a blue, low surface-brightness galaxy
(Roennback \& Bergval 1994).  The \HI\ extends slightly further
than the optical but is mostly unresolved by us
(Figure~\ref{fig:hispectra6}).  Roennback
\& Bergvall (1994) derive an optical inclination of $i$ =
63$^\circ$, and a P.A. = 43$^\circ$.  The \HI\ velocity field is
regular, with a P.A. similar to the optical.\\


\section{Discussion and Conclusions}

We have followed up, at higher spatial resolution with the ATCA, 
sixteen \HI\ sources in 6 southern GEMS groups 
that were previously detected by the Parkes telescope for which
the optical counterpart was not clearly identified (due to the
large beamsize of the Parkes telescope). 
The superior spatial resolution of the ATCA allows us to clearly
identify an optical counterpart in each case. 
We find a variety
of \HI\ structures and velocity fields.  These optical
counterparts are mostly low surface brightness late-type 
galaxies, some are clearly tidally interacting galaxies.

Three Parkes sources 
are clearly associated multiple optical counterparts.  
The Parkes source GEMS\_HCG90\_11 is resolved into \HI\ associated
with NGC 7204 and ESO 467-G002. There is no sign
of an interaction between them, however we note that NGC 7204
itself is a strongly interacting close pair of galaxies.   
Two Parkes sources in the NGC 7144 group appear to be associated
with interacting systems. 
Optical counterparts KTS 65 A\&B and ESO 236-G039 are in an interacting
system, which 
reveals a bridge of \HI\ within a common
\HI\ envelope. We suggest it will merge into a single gas-rich
galaxy on a $\sim$100 Myr timescale. The second Parkes source in this
group is associated with  ESO 236-G035 and ESO 236-G036. The \HI\
contours of both galaxies are roughly aligned with each other.  
Thus they may also form an interacting system with a relative velocity
of only 46 \kms\ between them. 
Interestingly, a friends-of-friends analysis indicates that the
NGC 7144 group has yet to form a coherent group structure and may
be an example of a group that is still collapsing 
(Brough et al.~2006).  

We find no
obvious cases for ram pressure stripping in the \HI\ structures. 
Most galaxies are at projected radii in the outskirts of the
groups, where any intragroup medium would be minimal. When 
irregular, extended \HI\ structures were seen they tended to be in
interacting systems. Kilborn et al. (2007) measured the \HI\
deficiency of the Parkes sources in our 6 groups and 10 other 
GEMS groups, concluding
that ram pressure stripping was not the dominant process in
removing \HI\ gas.

For two of the \HI\ sources (GEMS\_N3923\_11 and GEMS\_N5044\_18) we
have succeeded in identifying previously uncatalogued optical
counterparts. In both cases, the optical galaxies appear as faint, low
surface brightness objects on the DSS. The \HI\ velocity, and spatial
location, of these galaxies indicates that they are new members of the
NGC 3923 and NGC 5044 groups. We have also confirmed six new group
members in the NGC 3923, 5044 and 7144 groups.



Deeper \HI\ observations will no doubt 
reveal additional group members and hence further populate the
dwarf end of the group mass (and optical luminosity) function. 
The recent very deep \HI\ observations of Morganti et al. (2006)
indicate that the \HI\ mass function has a very long tail
to low \HI\ masses. This suggests 
that more dwarf galaxies in groups await discovery with future deep \HI\
observations.



\section*{Acknowledgments}


We thank Enno Middelberg for help with the ATCA \HI\ observing.
We also thank the Australian Research Council for financial support.

This research has made extensive use of the NASA/IPAC Extragalactic
Database (NED) which is operated by the Jet Propulsion Laboratory,
Caltech, under contract with the National Aeronautics and Space
Administration. The Digitized Sky Survey (DSS) was produced by the Space
Telescope Science Institute (STScI) and is based on photographic data
from the UK Schmidt Telescope, the Royal Observatory Edinburgh, the UK
Science and Engineering Research Council, and the Anglo-Australian
Observatory.

{}

\label{lastpage}

\end{document}